\pdfoutput=1
\documentclass[conference]{IEEEtran}

\usepackage{subcaption}
\usepackage{graphicx}
\usepackage{multirow}
\usepackage{amsmath}
\usepackage{color}
\usepackage{stfloats}
\usepackage{balance}
\usepackage{array} 
\usepackage{svg}

\begin{document}
\title{Low Latency Edge Classification GNN for Particle Trajectory Tracking on FPGAs}
\author{
\IEEEauthorblockN{Shi-Yu Huang\IEEEauthorrefmark{1}, Yun-Chen Yang\IEEEauthorrefmark{1}, Yu-Ru Su\IEEEauthorrefmark{1}, Bo-Cheng Lai\IEEEauthorrefmark{1}, Javier Duarte\IEEEauthorrefmark{2}, Scott Hauck\IEEEauthorrefmark{3}, Shih-Chieh Hsu\IEEEauthorrefmark{4}}, 
\IEEEauthorblockN{Jin-Xuan Hu\IEEEauthorrefmark{1}, Mark S. Neubauer\IEEEauthorrefmark{5}},
\IEEEauthorblockA{\IEEEauthorrefmark{1}Institute of Electronics, National Yang Ming Chiao Tung University, Taiwan}
\IEEEauthorblockA{\IEEEauthorrefmark{2}Department of Physics, University of California San Diego, 
 USA}
\IEEEauthorblockA{\IEEEauthorrefmark{3}Department of Electrical and Computer Engineering, University of Washington, USA}
\IEEEauthorblockA{\IEEEauthorrefmark{4}Department of Physics, University of Washington, USA}
\IEEEauthorblockA{\IEEEauthorrefmark{5}Department of Physics, University of Illinois at Urbana-Champaign, USA}
}

\par

\markboth{Journal of \LaTeX\ Class Files,~Vol.~14, No.~8, August~2015}
{Shell \MakeLowercase{\textit{et al.}}: Bare Demo of IEEEtran.cls for IEEE Journals}
\maketitle

\begin{abstract}



In-time particle trajectory reconstruction in the Large Hadron Collider is challenging due to the high collision rate and numerous particle hits. Using GNN (Graph Neural Network) on FPGA has enabled superior accuracy with flexible trajectory classification. However, existing GNN architectures have inefficient resource usage and insufficient parallelism for edge classification. This paper introduces a resource-efficient GNN architecture on FPGAs for low latency particle tracking. The modular architecture facilitates design scalability to support large graphs. Leveraging the geometric properties of hit detectors further reduces graph complexity and resource usage. Our results on Xilinx UltraScale+ VU9P demonstrate 1625x and 1574x performance improvement over CPU and GPU respectively.

\end{abstract}

\IEEEpeerreviewmaketitle

\section{Introduction}
\label{chapter:intro}



\par
Particle trajectory reconstruction in Large Hadron Collider (LHC) is a vital task for collision analysis, which requires accurate and in-time reconstruction to decide which collision events to read out \cite{ref27}. The existing reconstruction algorithms are based on Kalman filter \cite{ref15,ref16, ref17, ref18},  which are difficult to meet strict latency requirements due to the quadratically increasing complexity. The High-Luminosity LHC project \cite{ref28, ref29} aims to boost the instantaneous luminosity by 5x to 7x in 2027, even exacerbating the design challenges for trajectory reconstruction. Recent research shows that edge-classifying GNNs (Graph Neural Networks) achieve high accuracy in trajectory reconstruction and are scalable with the increased luminosity \cite{ref7,ref8,ref9,ref10,ref12,ref20}, making them a preferred solution for future collision analysis in LHC.


\par
One of the main concerns for GNNs to be implemented in the LHC system is the long processing latency with irregular data accesses. Recently proposed GNN accelerators \cite{ref22,ref23,ref35,ref36,ref37,ref38} are designed to focus on node data, such as Graph Convolutional Networks (GCNs) \cite{gcn} and GraphSAGE \cite{graphsage}. These cannot be applied directly to the edge classifying GNNs in trajectory reconstruction, which uses edge embedding and predicts the results on edges. Moreover, preprocessing on graphs is not fit for trajectory reconstruction which has relatively smaller but dynamic graph properties. State-of-the-art accelerators optimize their performance by reducing irregularity, such as rearranging processing patterns and order to better fit with their architectures \cite{ref23,ref38}, or monitor the utilization of processing elements (PEs) to balance the workload \cite{ref36}. These techniques are beneficial for static and large graphs, where stable graph characteristics can be reused. However, it is not efficient to spend long preprocessing time for one-time use on small graphs with dynamic features.


\par
In this paper, we propose an efficient architecture on FPGAs to support edge-classifying GNNs, meeting the timing requirement of LHC tasks. There are three novel contributions in the proposed architecture. First, a modular parallel architecture facilitates the design and scaling of the architecture with the size of the graphs. Second, efficient data allocation and buffer arrays considerably reduce memory conflicts during parallel data accesses. Third, exploitation of the geometry of collision events significantly lowers the graph irregularity by constraining the node connections, and thus increases processing parallelism.


\par
This work is implemented with the high-level-synthesis framework, hls4ml \cite{hls4ml_1,hls4ml_2}, which enables an automatic translation of machine learning models to FPGA designs. The experiments were performed on real collision graphs. The results on a Xilinx Virtex UltraScale+ VU9P FPGA show that the proposed GNN architecture achieves 1,625x and 1,574x speedup respectively compared with a Intel Xeon W-2125 CPU and an NVIDIA RTX2080 GPU. Section II of this paper discusses the background of particle tracking. Section III introduces the proposed GNN architecture. Section IV evaluates the performance and Section V concludes this work.


\section{Background}
\subsection{Large Hadron Collider System}
\begin{figure}[htb]  
	\centering 
	\includegraphics[width=0.5\textwidth]{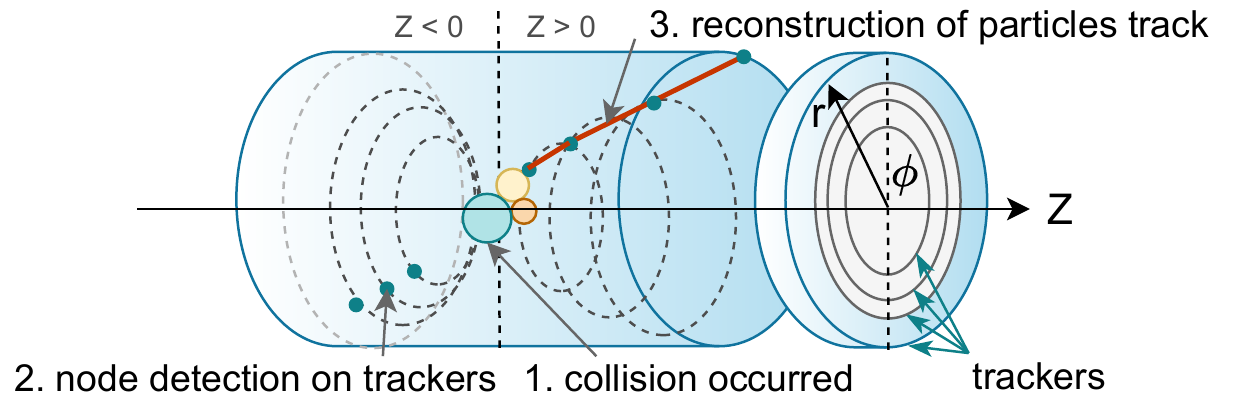} 
        \caption{Illustration of trajectory reconstruction} 
	\label{fig:LHC_Collision_Detector} 
\end{figure}

The Large Hadron Collider (LHC) is the largest and most powerful particle accelerator in the world \cite{ref28}. For high-energy particle physics collider experiments in the LHC, proton-proton collisions occur at a frequency of 40MHz and produce data at a rate of roughly 40 TB/s \cite{ref27}. After the collision, the trackers record the locations of particle detections (``hits") and transfer this information to the trigger systems. The trigger system will perform trajectory reconstruction to recognize which hits belong to the same particle, as shown in Fig.~\ref{fig:LHC_Collision_Detector}. The collision events are processed by 18 FPGAs in a multiplexed manner, where each FPGA needs to handle 2.22 million graphs per second (MGPS) \cite{cms2020}.

\par
Trackers are composed of cylindrical detecting layers \cite{ref28,ref20}. These layers are immersed in an axis-aligned magnetic field, and their geometry is naturally described by cylindrical coordinates. We focus on the innermost layers, a highly granular set of 4 barrel and 14 endcap layers \cite{ref39}.  

\subsection{GNN-based Algorithms for Track Reconstruction} \label{{GNN-based Algorithms for Track Reconstruction}}
The edge classifying GNN algorithm is based on interaction networks (IN) \cite{ref21}. IN is a physics-motivated GNN capable of analyzing objects and their relations. Hit information is embedded in the node feature, and trajectory segment information is embedded in the edge feature. The index set stores the sender and receiver node indexes of each edge. There are three types of functions in IN: Edgeblock, Aggregate, and Nodeblock. Functions in Edgeblock and Nodeblock are multi-layer perceptrons (MLPs) that re-embed edge and node features according to their input.  Aggregate accumulates edge features to their receiver nodes.

\subsection{Designs of GNN Accelerators on FPGAs}
\par
 Several studies have implemented GNNs on FPGAs for particle physics \cite{ref7,ref39,ref11}. \cite{ref7} focuses on jet tagging, which targets fully connected graphs and aims to predict features of the entire graph instead of each individual edge. While \cite{ref7} addresses the issue of irregular access with fully connected graph properties, it is not suitable for the LHC application. The graphs generated from the LHC consist of hundreds of nodes per graph. Utilizing the methods from \cite{ref7} may generate excessive unnecessary connections which may be tens to hundreds of times greater than the original graph, leading to a significant impact on processing time.
 


\section{A Low Latency GNN Architecture for Trajectory Reconstruction}\label{subsection:PF}

\begin{figure*}[ht]
      \begin{subfigure}{0.53\textwidth}
      \centering
      \includegraphics[width=\textwidth]{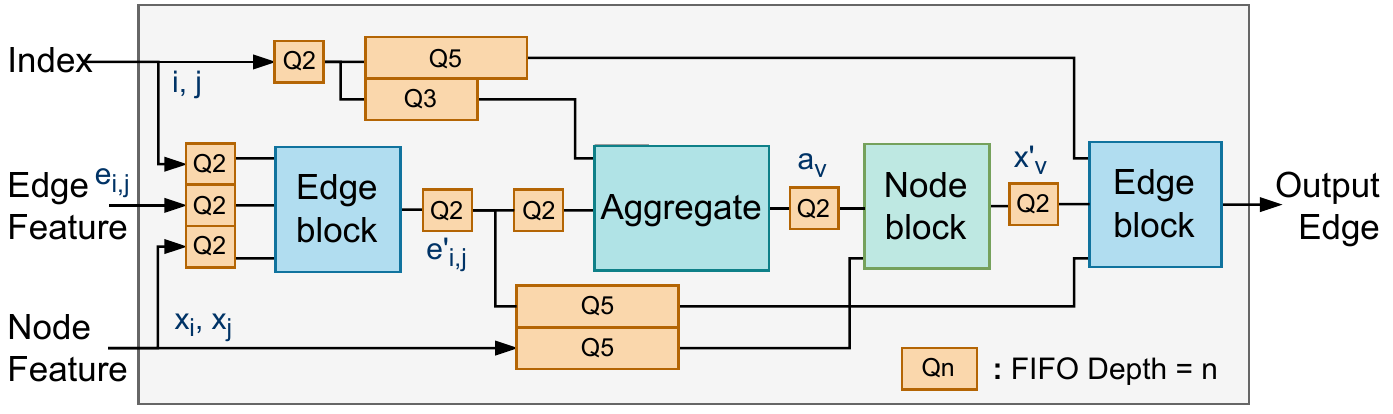}
      \caption{}
      \label{fig:overview_arc}
    \end{subfigure}
    \begin{subfigure}{.23\textwidth}
      \centering
      \includegraphics[width=\linewidth]{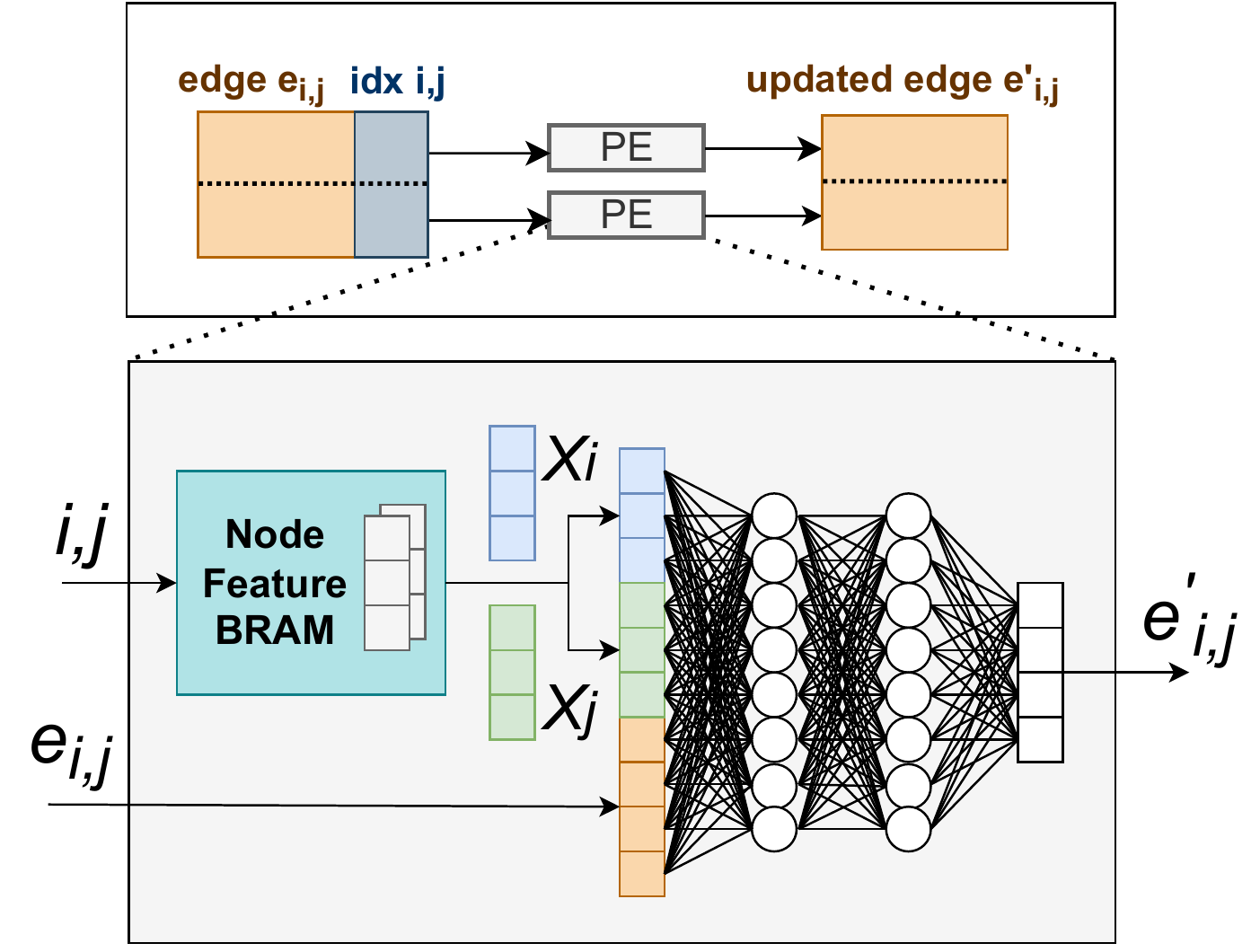}  
      \caption{}
      \label{fig:new_edgeblock}
     \end{subfigure}
    \begin{subfigure}{.23\textwidth}
      \centering
      \includegraphics[width=\linewidth]{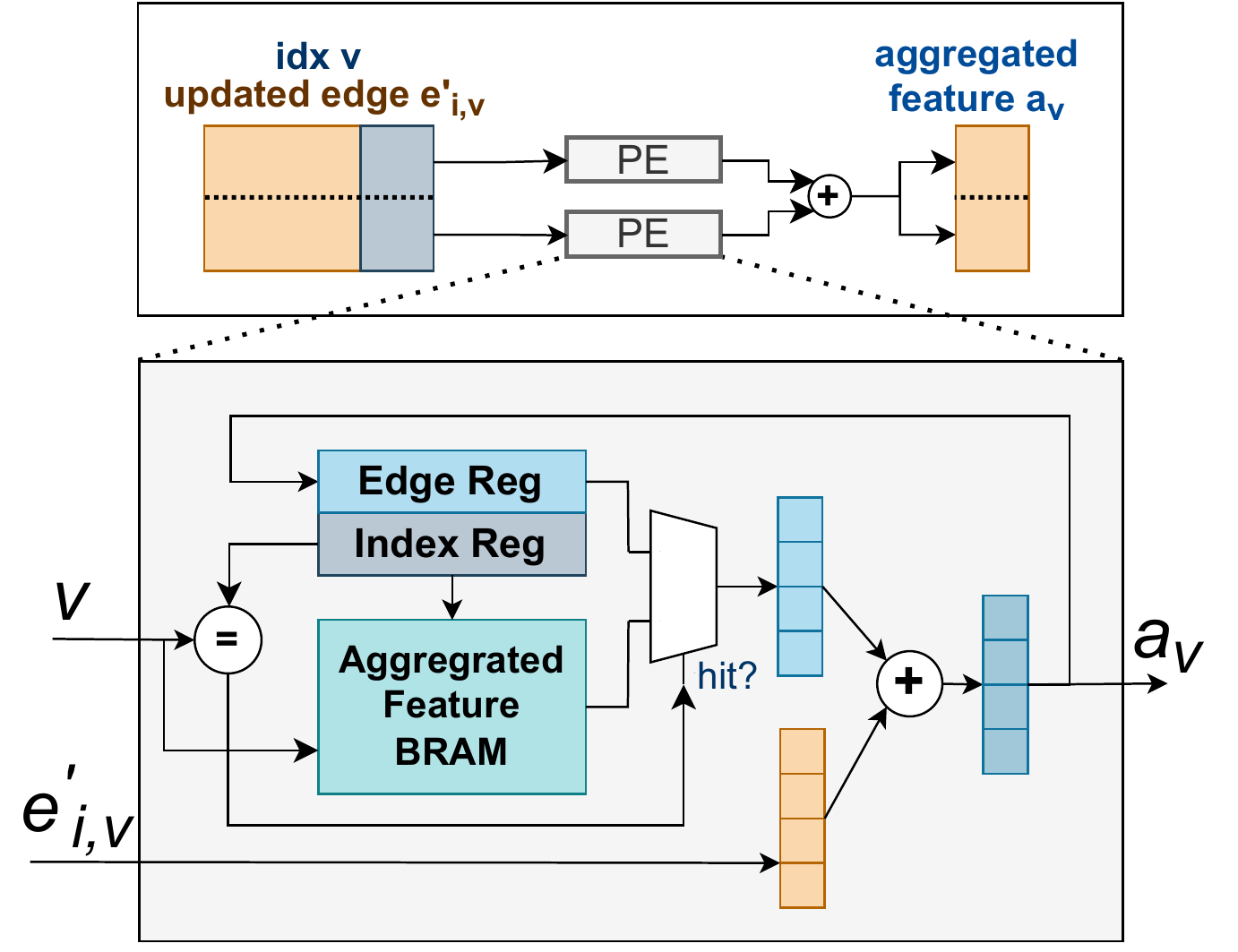}
      \caption{}
      \label{fig:new_aggregate}
    \end{subfigure}
    \caption{(a) Overview architecture of the system processing block. (b) An architecture of Edgeblock with two PEs. (c) The architecture of Aggregate with two FEs}
    \label{fig:test_sep}
\end{figure*}



\subsection{Overview Architecture}
Based on the GNN computation flow of IN in Section \ref{{GNN-based Algorithms for Track Reconstruction}}, the computation can be split into pipeline stages at the function level. We use the Vivado HLS dataflow architecture to implement the pipeline.  Fig.~\ref{fig:overview_arc} shows the proposed pipeline. We design a modularized parallel architecture for each function, including Edgeblock, Aggregate and Nodeblock. Each function is composed of several processing elements (PEs) as basic compute units. Between these functions, we insert FIFO buffers with different depths to ensure that data would not be stuck in the dataflow paths. With this architecture, users can configure the pipeline and scale the system throughput with the available resources on FPGAs.




\subsection{Modular Parallel Architecture}
\label{Modular Parallel Architecture}
The modular parallel architecture enables parallel processing of each function. The following introduces the design and optimizations of these functions.

\subsubsection{Edgeblock}
The Edgeblock computation involves accessing edge features, edge indexes, and connected node features. The access of node features changes dynamically according to the edge indexes. 
To address this irregularity, we added node arrays, which contain the features of all the nodes in the graph, into each PE to support concurrent accesses to node features. As shown in Fig.~\ref{fig:new_edgeblock}, during the computation, the edge features ($e_{i,j}$) and edge indexes (\textit{i}, \textit{j}) of each edge will be sent to PEs. In each PE, node features ($X_i$ and $X_j$) are accessed based on the edge indexes. After the above steps, multiplier engines in PEs will process the MLP computation and output the updated edge features. 


\subsubsection{Aggregate}
\par
The purpose of Aggregate is to send the updated edge features to their receiver nodes. During this process, the aggregate function first reads updated edge features and edge indexes. Based on the edge indexes, the aggregated edge features are accessed and added to the updated edge features, and then stored back to internal registers. The architecture of Aggregate PE is shown in Fig.~\ref{fig:new_aggregate}. 
After aggregating all updated edge features, the parallel adder tree accumulates the values of the same node indexes. With this architecture, multiple Aggregate PEs can process multiple edges simultaneously.



\subsubsection{Nodeblock}
The Nodeblock is used for re-embedding node features by the original and aggregated node features. The data access of the Nodeblock computation is more regular when compared with Edgeblock and Aggregate. For each node, Nodeblock collects node features and aggregated node features, and then uses MLPs to obtain updated node features.

\subsection{Exploiting Geometric Property of LHC Trackers} \label{Geometry-constrained}
While the architecture in the previous section successfully enables significant parallelism between processing elements (PEs), the individual memory in PEs costs considerable amount of BlockRAMs (BRAM) in an FPGA. Therefore, we propose a method that can reduce memory utilization and enable a more parallel architecture by taking advantage of the data properties of LHC detectors. 

\par
\begin{figure}[ht]
      \begin{subfigure}{.19\textwidth}
      \centering
      \includegraphics[width=\linewidth]{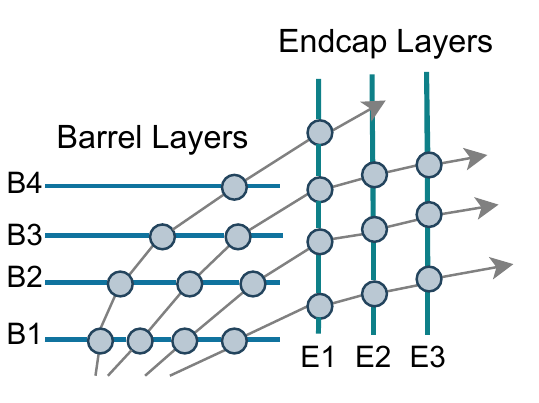} 
      \caption{}
      \label{fig:type}
    \end{subfigure}
    \begin{subfigure}{.17\textwidth}
      \centering
      \includegraphics[width=\linewidth]{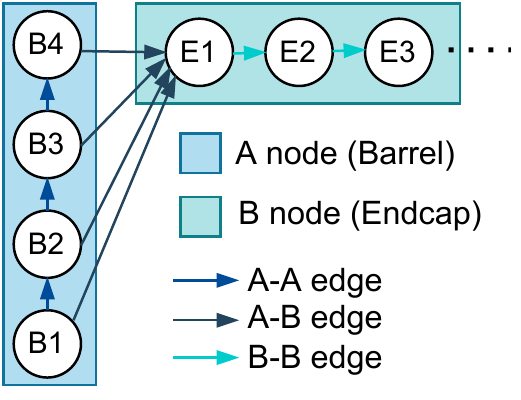}  
      \caption{}
      \label{fig:5_15}
     \end{subfigure}
    \begin{subfigure}{.1\textwidth}
      \centering
      \includegraphics[width=\linewidth]{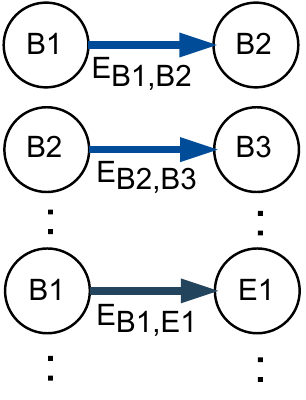}
      \caption{}
      \label{fig:connects}
    \end{subfigure}
    \caption{(a) Possible trajectories of particles, (b) Partition layers into two types of node group (c) Partition a graph into sub-graphs}
    \label{fig:test_sep2}
\end{figure}

In Section II-A, we introduced the architecture of the LHC particle trackers and how hit data is applied to the input graphs of GNNs. In the original graph constructed from LHC trackers, an edge from a node could connect to any other node in a graph. This assumption could cause excessive number of edges in the graph. Since the LHC trackers are composed of cylindrical layers surrounding the colliding beams, the particles have to pass through the inner tracker layer first and then move out, as shown in Fig.~\ref{fig:type}. These trajectories exhibit similar connection behaviors. For example, hits on the B1 layer only connect to the B2 layer or E1 layer. The relationship between hits and legal edges can be applied to a geometry-constrained graph. We reorganize the input graph structure by grouping hits based on their layer locations. We partition the graphs into 13 parallel sub-graphs, each containing only two node groups, as shown in Fig.~\ref{fig:5_15} and \ref{fig:connects}.

\par
With the geometry-constrained approach, an edge can only exist between specific node groups. This reduces the number of candidate nodes in node arrays used in Edgeblock PEs and Aggregate PEs, resulting in significant reduction of memory usage. Furthermore, since these 13 sub-graphs are independent of each other, they can be assigned to different PEs and computed in parallel. By relieving memory usage pressure and improving parallelism, the performance is greatly enhanced compared to the original design.
\section{Evaluation}

\subsection{Experiment Setup}
We implemented our design with Vivado HLS 2019.2 and loaded it onto a Xilinx Virtex UltraScale+ VU9P FPGA. The clock frequency of the design runs at 200 MHz. The resource utilization is from the v-synthesis report. The performance is based on  the simulation result of the generated HDL code. We use the metric of Million Graphs Per Second (MGPS) to measure the system throughput. 


\par
We evaluate our architecture with the TrackML dataset \cite{ref40} generated by CERN. 
All the data points are based on a fixed point format of 7 integer bits and 7 fractional bits. This is the same format used in \cite{ref39} to ensure acceptable accuracy. A system PE in the experiment contains an Edgeblock PE, an Aggregate PE, and a Nodeblock PE. The graph size of the dataset will be elaborated in Section \ref{Graph Size Analysis}. The performance of our proposed architecture will be evaluated in Sections \ref{Scalability Analysis of Modularized Parallel Architecture} to \ref{Advanced Analysis of Geometry-constrained Graphs}. In Section \ref{comparison}, we compare the performance of our architecture with the CPU, GPU, and prior FPGA designs.

 \subsection{Supporting In-time Graph Processing of Collision Events} \label{Graph Size Analysis}
The ultimate goal of this work is to perform in-time trajectory reconstruction based on the graphs generated from LHC collision events. Input graphs are prepared based on the same flow as the prior work \cite{ref39}. Each graph is divided into two sectors based on the position $z$ of hits. We use the graph size that can cover 95 percentile of collision events as the nominal size, which contains 739 nodes with 1252 edges. According to \cite{cms2020}, these graphs  should be computed at the throughput higher than 2.22 MGPS. 

\par
Table~\ref{table:5_5_re_pe} compares the three proposed architectures. The architecture MPA represents the Modular Parallel Architecture introduced in Section \ref{Modular Parallel Architecture}. MPA\textsubscript{geo} and MPA\textsubscript{geo\_rsrc} are the extended designs of MPA with the proposed techniques of geometry-constrained optimization and data-aware resource allocation respectively. The designs of MPA\textsubscript{geo} and MPA\textsubscript{geo\_rsrc} will be elaborated in Section \ref{Influence of Geometry-constrained Implementation} and \ref{Advanced Analysis of Geometry-constrained Graphs}. Latency measures the time from input graph to the output result. The design can take a new input in every Interval time and attain throughput in MGPS. As shown in Table~\ref{table:5_5_re_pe}, the proposed MPA\textsubscript{geo\_rsrc} meets the LHC requirement by supporting graphs of 739 nodes with 1252 edges at throughput of 3.225 MGPS.  

\begin{table}[htb]
    \centering 
    \renewcommand\arraystretch{1.3}
	\caption{Performance of the proposed architectures} 
        \label{table:5_5_re_pe}
        \resizebox{.48\textwidth}{!}{
        \begin{tabular}{ c | c c c c c}
        \hline
           Architectures & Latency($\mu$s) & Interval($\mu$s) & Throughput(MGPS) \\ 
        \hline
            MPA & 3.165 & 0.48 & 2.083 \\ 
        \hline
            MPA$_\text{geo}$ & 2.69 & 0.425 & 2.352 \\ 
        \hline
            MPA$_\text{geo\_rsrc}$ & 2.07 & 0.31 & 3.225 \\ 
        \hline
        \end{tabular}
}
\end{table}


\subsection{Scalability of MPA (Modular Parallel Architecture)} \label{Scalability Analysis of Modularized Parallel Architecture}
\begin{figure}[h]
    \centering 
    \includegraphics[width=.4\textwidth]{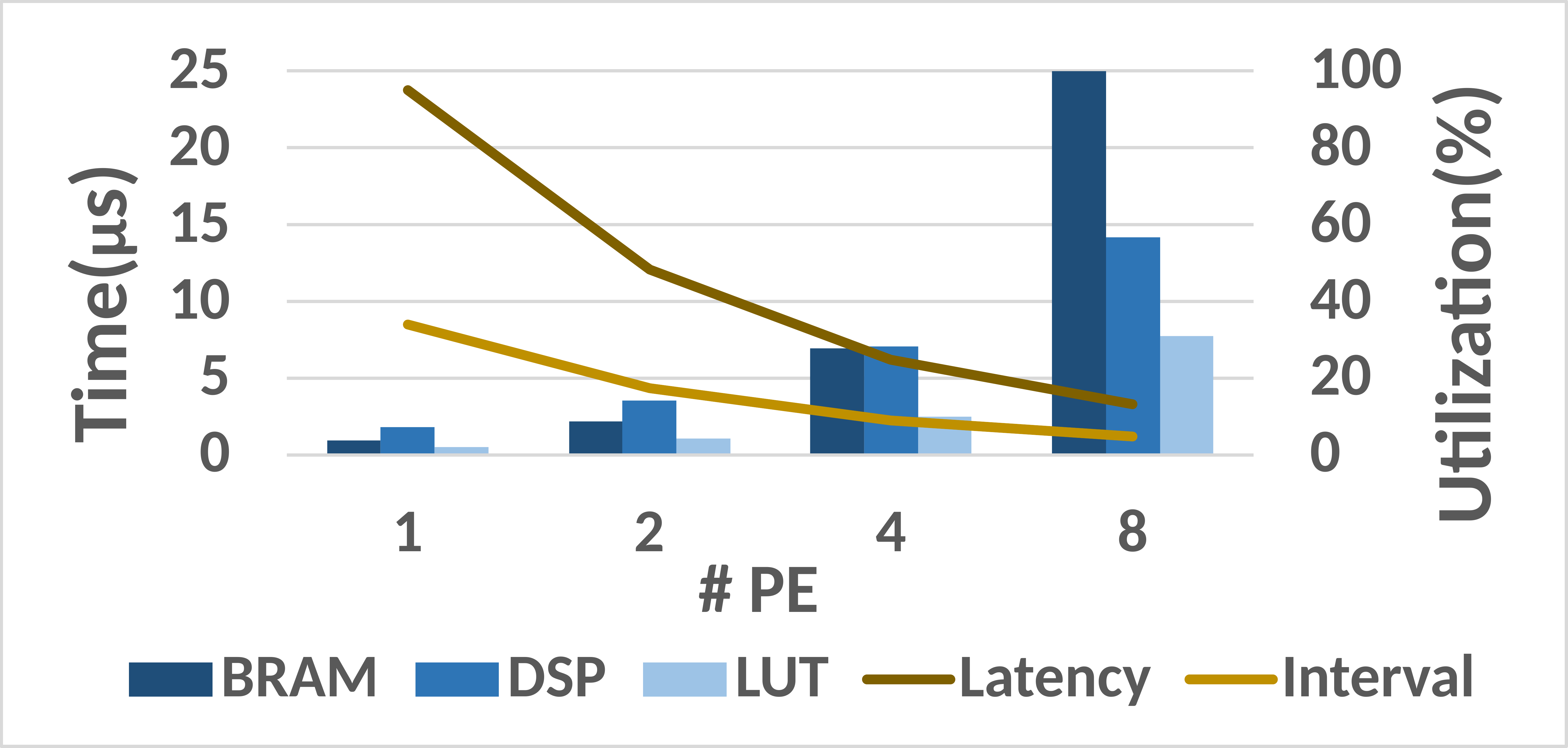}
    \caption{Scalability of the modular architecture} 
    \label{fig:scale_fuse} 
\end{figure}

In Section III-B, we introduced the MPA architecture. The processing throughput of the architecture can scale by deploying more PEs. To evaluate the scalability of MPA, Fig.~\ref{fig:scale_fuse} illustrates the performance and resource utilization of MPA from one PE to eight PEs. The results show the latency and interval can be reduced by deploying more PEs. However, when scaling up the number of PEs, BRAMs will become the limiting factor of FPGA resources.  


\subsection{MPA with Geometry-constrained Optimization}
\label{Influence of Geometry-constrained Implementation}
By taking advantage of the geometry-constrained property described in Section III-C, MPA\textsubscript{geo} not only relieves the constraints of the range of node in each PE, but also makes node groups independent to others. MPA\textsubscript{geo} alleviates the resource demand of BRAM and allows the deployment of more PEs for greater processing parallelism. There are 11 node groups and 13 edge groups. We allocate one PE for each of these groups and result in a total of 11 Nodeblock PEs, 13 Edgeblock PEs, and 13 Aggregate PEs. As shown in Table~\ref{table:5_5_re_pe}, MPA\textsubscript{geo} achieves 13\% improvement in throughput compared to the original MPA architecture.

\subsection{Data-aware Resource Allocation}
\label{Advanced Analysis of Geometry-constrained Graphs}
We further analyze the distribution of graph sizes in the dataset and propose the design MPA\textsubscript{geo\_rsrc} which applies data-aware resource allocation. 
After applying the geometry-constrained property, the number of nodes are not evenly distributed across different layers. The barrel layers (B1 to B4) contain more nodes and connections than endcap layers (E1 to E7). To address this issue, we propose the design MPA\textsubscript{geo\_rsrc} which classifies the node groups into two types. As shown in Fig.~\ref{fig:5_15}, the layers B1 to B4 contain relatively more nodes and belong to type A, while layers E1 to E7 with fewer nodes are assigned to type B. We will assign two PEs to process each node group of type A, and one PE to handle each node group in type B. For the edge groups, we apply the same allocation principle as for node groups.




\begin{table}[htb]
    \centering 
    \renewcommand\arraystretch{1.3}
	\caption{Allocate PEs based on different sizes of data} 
        \label{table:dis}
            \begin{tabular}{c|cc|ccc}
            \hline
            & \multicolumn{2}{c|}{Node} & \multicolumn{3}{c}{Edge} \\ \hline
            & A            & B          & A-A    & A-B    & B-B    \\ \hline
            \#data& 138          & 62         & 277    & 77     & 87     \\ \hline
            \#PE  & 2            & 1          & 4      & 1      & 1      \\ \hline
        \end{tabular}	
\end{table}

\par

\subsection{Comparison with Previous Designs}
\label{comparison}
\subsubsection{Comparison with Previous GNN Trajectory Reconstruction on FPGA}
 There are two architectures in the previous work \cite{ref39} of GNN for trajectory reconstruction on FPGA. The throughput-optimized design (ThrpOpt) focuses on attaining high throughput, but would reduce the graph size it can handle. The resource-optimized design (RsrcOpt) aims to accommodate large graphs, but would suffer from low throughput. Table~\ref{table:baseline compared} compares the performance between these two architectures and our proposed architecture. The platform of all the three architectures is XCVU9P, and the frequency is 200 MHz. ThrpOpt design can achieve a higher throughput of 200 MGPS, but can only handle small graphs of 28 nodes with 56 edges. RsrcOpt architecture can accommodate large graphs of 448 nodes with 896 edges, but with lower throughput than the ThrpOpt design. Our proposed MPA\textsubscript{geo\_resrc} can handle the largest graph (739 nodes with 1252 edges) among all the designs, and attains higher throughput than the RsrcOpt design. 

\begin{table}[h]
\centering
\renewcommand\arraystretch{1.3}
\caption{Comparison with previous FPGA designs}
\resizebox{.48\textwidth}{!}{
\begin{tabular}[htb]{ c|c c c }
    \hline
     & ThrpOpt \cite{ref39} & RsrcOpt \cite{ref39} & MPA\textsubscript{geo\_rsrc} (proposed) \\
     \hline
    Graph Size & 28 nodes/56 edges & 448 nodes/896 edges& 739 nodes/1252 edges \\
    \hline 
    Throughput & 200 MGPS & 1.14 MGPS & 3.17 MGPS\\
    \hline
\end{tabular}}
\label{table:baseline compared}
\end{table}

\subsubsection{Comparison with CPU and GPU}


We execute the same particle-tracking GNN algorithm on an Intel(R)Xeon(R) W-2125 CPU and an NVIDIA GeForce RTX 2080 Ti (CUDA 10.2) based on PyTorch (1.11.0) and the PyTorch Geometric 2.0.4 framework.
We ran 1000 graphs on each platform. Each graph contains 739 nodes and 1252 edges. Table~\ref{cpugpu} shows the details of experiment and normalized throughput. Our proposed design on FPGA achieved significantly higher throughput of 1,625x and 1,574x when compared with CPU and GPU respectively.


\begin{table}[htb]
\centering
\renewcommand\arraystretch{1.3}
\caption{Comparison with CPU and GPU}
\resizebox{.48\textwidth}{!}{
\begin{tabular}[htb]{ c| c c c }
    \hline
     & CPU & GPU & FPGA \\
     \hline
    Platform & Intel(R) Xeon(R) & NVIDIA GeForce & XCVU9P \\
    & W-2125  & RTX 2080 Ti & \\
    \hline
    Compute Unit & 4.00 GHz@8 cores & 1.63 GHz@4352 cores & 200 MHz \\
    \hline
    Technology & 14 nm & 12 nm & 14 nm \\
    \hline
    Normalized Thrp. & 1 & 1.03 & 1625 \\
    \hline
\end{tabular}}
\label{cpugpu}
\end{table}
\section{Conclusion}
We propose a novel architecture for particle-tracking GNNs on FPGAs.  By utilizing LHC detector geometry, our design reduces graph complexity and FPGA resource requirements. The modular architecture of processing units and buffers also efficiently handle the irregular data access patterns and facilitate design scalability to support large graphs while attaining high parallelism and computation throughput. Experiment results show that our design achieves 1,625x speedup compared to the CPU, and 1,574x speedup compared to the GPU.


\section{Acknowledgments}
Huang, Yang, Su, Lai and Hu are supported by National Science and Technology Council grant 111-2221-E-A49-092-MY3. Duarte, Hauck, Hsu and Neubauer are supported by National Science Foundation (NSF) grants No. 2117997.
\par

\balance
\bibliographystyle{IEEEtran}
\bibliography{ref}

\end{document}